\documentclass[%
reprint,
superscriptaddress,
amsmath,amssymb,
aps,
prd,
]{revtex4-2}

\usepackage{graphicx}
\usepackage{dcolumn}
\usepackage{bm}
\usepackage{xspace}
\usepackage{xcolor}
\usepackage{mathtools} 
\usepackage{braket}
\usepackage{hyperref}
\definecolor{darkblue}{rgb}{0.18,0.19,0.57}
\hypersetup{
    colorlinks,
    linkcolor=darkblue,
    citecolor=darkblue,
    urlcolor=darkblue
}

\definecolor{darkgreen}{rgb}{0.0, 0.5, 0.0}

\newcommand{\nn}{\nonumber\\}

\newcommand{\f}[1]{\mbox{\boldmath$#1$}}
\newcommand{\fk}[1]{\mbox{\boldmath$\scriptstyle#1$}}

\newcommand{\bea}{\begin{eqnarray}}
\newcommand{\ea}{\end{eqnarray}}
\newcommand{\eea}{\end{eqnarray}}
\newcommand{\ord}{{\cal O}}

\begin{document}

\preprint{APS/123-QED}

\title{Quantum radiation in dielectric media with dispersion and dissipation} 

\author{Sascha Lang} \email{s.lang@hzdr.de} 

\affiliation{%
 Helmholtz-Zentrum Dresden-Rossendorf, Bautzner Landstra{\ss}e 400, 01328 Dresden, Germany
}
\affiliation{%
 Fakult\"at f\"ur Physik, Universit\"at Duisburg-Essen, Lotharstra{\ss}e 1, 47057 Duisburg, Germany
}
\author{Ralf Sch\"utzhold}%
\affiliation{%
 Helmholtz-Zentrum Dresden-Rossendorf, Bautzner Landstra{\ss}e 400, 01328 Dresden, Germany
}%
\affiliation{%
 Institut f\"ur Theoretische Physik, Technische Universit\"at Dresden, 01062 Dresden, Germany
}
\affiliation{%
 Fakult\"at f\"ur Physik, Universit\"at Duisburg-Essen, Lotharstra{\ss}e 1, 47057 Duisburg, Germany
}
\author{William G. Unruh}
\affiliation{
 Department of Physics and Astronomy, University of British Columbia, Vancouver V6T 1Z1, Canada
}%
\affiliation{
 	Hagler Institute for Advanced Study, Institute for Quantum Science and Engineering, Texas A\&M University, College Station, Texas 77843-4242, USA
}%

\date{\today}

\begin{abstract}
By a generalization of the Hopfield model, we construct a microscopic 
Lagrangian describing a dielectric medium with dispersion and dissipation. 
This facilitates a well-defined and unambiguous {\em ab initio} 
treatment of quantum electrodynamics in such media, 
even in time-dependent backgrounds. 
As an example, we calculate the number of photons created by switching on and 
off dissipation in dependence on the temporal switching function. 
This effect may be stronger than quantum radiation produced by variations 
of the refractive index $\Delta n(t)$ since the latter are 
typically very small and yield photon numbers 
of order $(\Delta n)^2$.
As another difference, we find that the partner particles of the created 
medium photons are not other medium photons but excitations of the
environment field causing the dissipation (which is switched on and off). 
%
\end{abstract}

\maketitle


\section{Introduction} 

One of the most striking consequences of quantum field theory is the non-trivial 
nature of the vacuum or ground state. 
Even in this lowest-energy state, fields do not vanish identically, but are 
permanently fluctuating. 
These quantum vacuum fluctuations cause many well-known effects such as spontaneous 
emission \cite{Weisskopf_1930}, the Casimir effect \cite{Casimir_1948}, 
or the Lamb shift \cite{Lamb_1947, Karplus_1951}. 
Another fascinating consequence is the phenomenon of quantum radiation, where these 
fluctuations are converted into real particles by suitable
external conditions, which would have no effect on the classical vacuum 
(with all fields vanishing identically).
Examples include Hawking radiation \cite{HawkingNature_1974, HawkingComm_1975}, 
the dynamical Casimir effect \cite{Moore_1970},
and cosmological particle creation \cite{Schroedinger_1939, Parker_1968}; 
but also time (or even space-time) dependent variations of the refractive index 
in dielectric media (or waveguides), 
where the latter can display interesting analogies 
\cite{Unruh_1981, Visser_1998, Barcelo_2011}
to the former ones, see also 
Refs.~\cite{Leonhardt_2002, Unruh_2003, SchuetzholdUnruh_2005, 
Philbin_2008, Nation_2009, Belgiorno_2010, Schuetzhold_2011, Liberati_2012,
  Finazzi_2013,  Belgiorno_2015, Jacquet_2015, Linder_2016, 
 Jacquet_2018, Drori_2019, Jacquet_2020}, 
\cite{Johansson_2009, Johansson_2010, Wilson_2011, Laetheenmaeki_2013} and 
\cite{Westernberg_2014, Tian_2017, Lang_2019}, respectively.

In a notably simplified approach, aspects of quantum radiation 
can be studied by neglecting medium properties such as dispersion and dissipation. 
Going beyond this simple picture, there has been considerable work regarding 
the effects of dispersion, see, e.g., 
\cite{Finazzi_2013, Belgiorno_2014,  Belgiorno_2015, Jacquet_2015, Linder_2016, 
 Jacquet_2018, Jacquet_2020}. 
However, in the vast majority of publications, quantum radiation has been 
considered in absence of dissipation, with a few 
exceptions including \cite{Adamek_2008, Parentani_2008, Robertson_2015}.
One of the main reasons lies in the intrinsic difficulty of treating 
dissipation correctly, especially regarding quantum fluctuations under 
non-trivial external conditions.

There are basically two main approaches for adding dissipation to the 
well-established theory of non-dissipative dielectrics discussed in, e.g.,  
\cite{Hopfield_1958, Huttner_1991, Belgiorno_2016}.
In a top-down approach, one starts with the phenomenological properties of a 
given medium such as the complex dielectric permittivity $\varepsilon(\omega)$ 
and then constructs the corresponding quantum field operators by demanding 
consistency conditions, see, e.g., \cite{Gruner_1995, Gruner_1996, Scheel_1998, BuhmannScheel_2008}.
The alternative bottom-up approach, on the other hand, is based on microscopic models, 
which allow for deriving the associated medium properties such as $\varepsilon(\omega)$.
For simple cases, such as stationary and homogeneous media, it is possible to show the 
equivalence of these two approaches via the Huttner-Barnett formalism 
\cite{Huttner_1992Letter, Huttner_1992, Dutra_1998} 
based on an exact Fano diagonalization \cite{Fano_1956, Fano_1961, Costa_2000}.  
However, extending this formalism to more general cases such as 
temporally and possibly even spatio-temporally varying media is quite involved.
Thus, even though the phenomenological approach has the obvious advantage 
to account for media with very general $\varepsilon(\omega)$, it has the 
drawback of potential ambiguities, especially in time-dependent scenarios.

A related issue is the explicit calculation of observables 
(e.g., the number of created photons) which typically requires 
certain approximations. 
In order to describe dissipative media, several microscopic approaches
employ a Markov-type approximation (e.g., in Weisskopf-Wigner theory) 
which neglects the memory of the environment, see also \cite{Knoell_1992}.
Especially in time-dependent scenarios, the justification and 
applicability of such an approximation must be scrutinized in 
order to avoid inconsistencies. 

In the following, we propose and study an explicit microscopic model 
(bottom-up approach) for a 1+1 dimensional dielectric medium 
including dispersion and dissipation, which does not require 
any Markov-type approximations and has well defined in 
and out states. 
The goal is an {\em ab inito} treatment of quantum radiation 
without ambiguities and additional assumptions. 
Using this approach, we study quantum radiation emerging from time-dependent 
variations (switching on and off) in the coupling between a medium  
and its environment, see also \cite{Ciuti_2005, Liberato_2007, Auer_2012, Liberato_2017}. 

\section{The Model}\label{sec:classicalModel}

We consider the following Lagrangian 
\begin{equation}
L = L_{A} + L_{\Psi} + L_{A\Psi} + L_{\Phi} + L_{\Psi\Phi}
\,,
\label{eq:classicalLagrangianSum}
\end{equation}
where $L_{A}$ describes the electromagnetic vector potential $A(t,x)$ 
in 1+1 dimensions ($\hbar=c=1$)
\begin{equation}
L_{A}    = \frac{1}{2} \int dx \,
\left\lbrace \left[\partial_t A(t,x)\right]^2 - \left[\partial_x A(t,x)\right]^2
\right\rbrace
\,.
\label{eq:LA}
\end{equation}
As usual in the Hopfield model, the polarization of the medium 
is included by adding harmonic oscillators $\Psi(t,x)$ with 
resonance frequency $\Omega>0$ to all points $x$ 
of the dielectric
\begin{equation}
L_{\Psi}   = \frac{1}{2} \int dx \,
\left\lbrace \left[\partial_t \Psi(t,x)\right]^2  -  \Omega^2 \Psi^2(t,x)
\right\rbrace
\,,
\label{eq:LPsi}
\end{equation}
and coupling them to the electric field $E=-\partial_t A$ via 
\begin{equation}
L_{A\Psi} = -g \int dx \,
\Psi(t,x)\partial_t A(t,x) 
\,,
\label{eq:LAPsi}
\end{equation}
with the coupling strength $g$. 

The above terms $L_{A} + L_{\Psi} + L_{A\Psi}$ represent the usual 
Hopfield model \cite{Hopfield_1958, Huttner_1991, Belgiorno_2016, Linder_2016}. 
In order to include dissipation, we introduce an additional field 
$\Phi(t,x,y)$ which can exchange energy with 
the medium and propagates in a perpendicular ($y$) direction 
\begin{equation}
L_{\Phi}   = \frac{1}{2} \int dx\, dy\,
\left\lbrace\left[\partial_t \Phi(t,x,y)\right]^2 
  -  \left[\partial_y \Phi(t,x,y)\right]^2\right\rbrace
\,.
\label{eq:LPhi}
\end{equation}
This field is coupled to the medium $\Psi(t,x)$ in the same way 
as the electromagnetic field $A(t,x)$, but with a coupling strength $G$ 
\begin{equation}
L_{\Psi\Phi} = -G \int dx\,\Psi(t,x)\partial_{t}\Phi(t,x,y=0) 
\,,
\label{eq:LPhiPsi}
\end{equation}
where we assume the medium to be located along the $y=0$ line. 

In principle, this model holds for media with general time-dependent parameters 
$\Omega$, $g$ and $G$, and can even be generalized to fully space-time dependent settings.

\section{Equations of Motion}

In order to show that the above model~\eqref{eq:classicalLagrangianSum}
does indeed feature the dynamics expected for a dissipative medium, 
let us study the associated Euler-Lagrange equations. 
For the electromagnetic field $A(t,x)$, we obtain the same form as in the 
usual Hopfield model 
\begin{equation}
\left[\partial_t^2 -  \partial_x^2\right]  A(t,x) 
=  \partial_t  \left[ g \Psi(t,x)\right]
\,,
\label{eq:eulerLagrangeEquationA}
\end{equation}
but the medium field $\Psi(t,x)$ acquires an additional term 
\begin{equation}
\left[\partial_t^2 + \Omega^2\right] \Psi(t,x) =   
-g \partial_t  A(t,x)
-G \partial_t \Phi(t,x,0)
\,.
\label{eq:eulerLagrangeEquationPsi}
\end{equation}
Finally, the environment field $\Phi(t,x,y)$ evolves according to  
\begin{equation}
\left[\partial_t^2 - \partial_y^2\right] \Phi(t,x,y)
= \partial_t\left[G\Psi(t,x)\right] \delta(y)
\,,
\label{eq:eulerLagrangeEquationPhi}
\end{equation}
where we have written all 
equations is such a way that they equally 
hold for time-dependent $\Omega(t)$, $g(t)$ and $G(t)$. 

\subsection{Dispersion relation}\label{sec:Dispersion}

Considering constant parameters
($\Omega$, $g$ and $G$) for the moment, 
we may solve Eq.~\eqref{eq:eulerLagrangeEquationPhi} via the retarded 
Green's function and arrive at 
\begin{equation}
\Phi(t,x,y) = 
\Phi_0(t,x,y) + \frac{G}{2}\Psi(t - \vert y \vert, x) \,,
\label{eq:PhiFormalSolution}
\end{equation}
where $\Phi_0(t,x,y)$ denotes the homogeneous solution 
of Eq.~\eqref{eq:eulerLagrangeEquationPhi}, i.e., of 
$\left[\partial_t^2 - \partial_y^2\right] \Phi_0(t,x,y) = 0$. 
Since we have used the retarded Green's function 
(with the retarded time argument $t-|y|$), this solution 
$\Phi_0(t,x,y)$ describes the environment field originating 
from $\mathcal{I}^-$ (i.e., $t\to-\infty$ and $y\to\pm\infty$) 
before interacting with the medium at $y=0$. 

Inserting this solution back into Eq.~\eqref{eq:eulerLagrangeEquationPsi},
we get a driven and damped oscillator at each position $x$ 
\begin{eqnarray}
\left[\partial_t^2 + \frac{G^2}{2}\,\partial_t + \Omega^2\right]\Psi(t,x) 
&=& - G \partial_t \Phi_0(t,x,y=0)
\nonumber\\
&&
-g \partial_t A(t,x)
\,,
\label{eq:damped-oscillator}
\end{eqnarray}
where we can read off the damping factor $\Gamma=G^2/4$ 
of the medium.  
By finally combining Eqs.~\eqref{eq:eulerLagrangeEquationA}
and \eqref{eq:damped-oscillator}, we find (for constant 
$\Omega$, $g$ and $G$) 
\bea
\label{eq:A-from-Phi}
\left[
\left(\partial_t^2+\frac{G^2}{2}\,\partial_t+\Omega^2\right)
\left(\partial_t^2-\partial_x^2\right)
+g^2\partial_t^2 
\right]
A(t,x)
=
\nn
-gG\partial_t^2 \Phi_0(t,x,y=0)\,.\quad
\ea
The environment field on the right-hand side 
constitutes the classical counterpart of 
the quantum noise term required by the fluctuation-dissipation theorem 
while the differential operator on the left-hand side yields 
the dispersion relation 
\bea
k^2
=
\omega^2\left(1+\frac{g^2}{\Omega^2-i\omega G^2/2-\omega^2}\right)
=
\omega^2\varepsilon(\omega)
\,,
\ea
which turns into  a standard textbook expression 
(see, e.g., Sec. 11.3 of Ref. \cite{IbachLueth_2003}) for dissipative dielectric media 
after some minor rescaling of system parameters. 

\begin{figure}[htbp]
\centering
\begin{minipage}{\linewidth}
\includegraphics[width=0.95\linewidth]{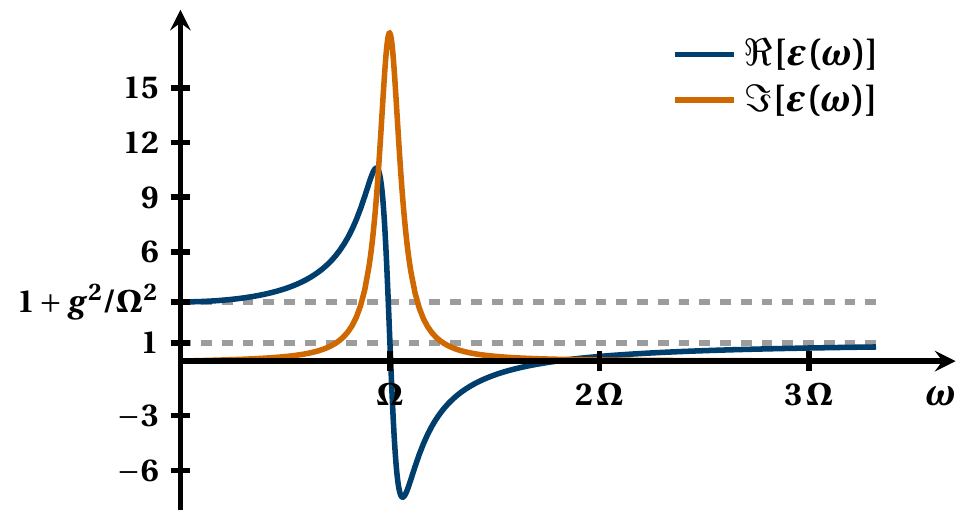}
\end{minipage}
\caption{Real and imaginary parts of the dielectric permittivity 
$\varepsilon(\omega)$ 
for exemplary parameters $g = 3 \Omega/2$ and 
$G = \sqrt{\Omega}/2$, i.e., 
in the underdamped regime of $G^2 < 4 \Omega$. 
}
\label{fig:ReAndImOfEpsilon}
\end{figure}

From the corresponding dielectric permittivity $\varepsilon(\omega)$ 
illustrated in Fig.~\ref{fig:ReAndImOfEpsilon}, 
we obtain the effective refractive index $n = \sqrt{1+g^2/\Omega^2}$ 
for small photon frequencies $\omega$. 
The imaginary part 
\bea
\label{eq:A-Field-Damping}
\Im[\varepsilon(\omega)]=2\Gamma\,\frac{n^2-1}{\Omega^2}\,\omega+\ord(\omega^2)
\ea
is closely linked to the damping exponent of solutions 
\bea
\label{eq:dampedA-solution}
A(t,x) \propto \exp{\left\{ i \omega \sqrt{\varepsilon(\omega)} x - i  \omega t \right\}}
\ea
which oscillate  at frequencies $\omega > 0$ where 
\bea
\label{eq:damping-strength}
\Im\left\{\sqrt{\varepsilon(\omega)}\right\}  
=  
\Gamma\,\frac{n^2-1}{n \Omega^2}\,\omega + \mathcal{O}\left(w^2\right)
\,.
\ea
Note that this quantity is related to but not identical 
with the intrinsic damping $\Gamma = G^2/4$ of the oscillators $\Psi(t,x)$.

\section{Quantization}

As an advantage of our microscopic model~\eqref{eq:classicalLagrangianSum}, we may now 
derive the corresponding quantum field operators $\hat{A}$, $\hat{\Psi}$, 
and $\hat{\Phi}$ in an unambiguous manner which consistently takes the quantum fluctuations 
of all fields into account. 
The origin of those fluctuations depends on the time-dependence of the coupling parameters
$g(t)$ and $G(t)$. 
If both adopt finite values for a sufficiently long time, extending far into the past, 
all quantum fluctuations can ultimately be traced back to 
vacuum fluctuations of the environment field $\hat\Phi_0$.
However, if that field is initially decoupled from the 
remaining ones $\hat A$ and $\hat\Psi$ (the case we shall consider later), 
the $\hat A$ and $\hat\Psi$ fields will bring in their own initial quantum fluctuations. 
Then, if $G(t)$ is switched on, all these  fluctuations become mixed -- 
which gives rise to particle creation.

\subsection{Environment Field $\f{\Phi_0}$}

Let us first consider the impact of the environment field. 
In the general case of time-dependent parameters $\Omega(t)$, $g(t)$ and $G(t)$, 
the equations of motion~\eqref{eq:eulerLagrangeEquationA}, 
\eqref{eq:eulerLagrangeEquationPsi} and 
\eqref{eq:eulerLagrangeEquationPhi} 
can be decoupled analogous to the scenario of a static medium. 
For example, the solution of Eq.~\eqref{eq:eulerLagrangeEquationPhi} for time-dependent 
$G(t)$ has the same form as in Eq.~\eqref{eq:PhiFormalSolution} but 
with the modified argument $G(t-|y|)$.
Inserting this solution back into Eqs.~\eqref{eq:eulerLagrangeEquationA}
and \eqref{eq:eulerLagrangeEquationPsi}, we may decouple these two 
second-order equations into one fourth-order equation for the medium field 
\bea
\label{eq:decouple-Psi}
\left\lbrace 
\left[\partial_t^2 - \partial_x^2\right] \frac{1}{g}
\left[\partial_t^2 + \frac{1}{2} G\partial_tG + \Omega^2\right]
+ \partial_t^2 g
\right\rbrace\hat{\Psi}(t,x)
=
\nn
-\left[\partial_t^2 - \partial_x^2\right] \frac{G}{g} \partial_t \hat{\Phi}_0(t,x,y=0)
\,,\quad
\ea
where we have omitted the arguments $t$ of all system parameters 
$\Omega$, $g$ and $G$ to enhance readability.
After solving Eq.~\eqref{eq:decouple-Psi}, the corresponding 
electromagnetic field $\hat{A}(t,x)$ can finally be obtained via integration of 
Eq.~\eqref{eq:eulerLagrangeEquationPsi}.

If the initial $G(t \to -\infty)$ is non-vanishing, 
the homogeneous solutions of the above equation 
decay with time and thus only the inhomogeneous solution 
stemming from the source term on the right-hand side survives. 
In other words, the initial quantum fluctuations of the fields $\hat A$ and $\hat\Psi$ are 
transferred (i.e., lost) to the environment and all the remaining quantum fluctuations stem from 
the primordial fluctuations of $\hat{\Phi}_0$.

As explained above, this homogeneous solution $\hat\Phi_0(t,x,y)$ constitutes 
a free scalar field in two spatial dimensions, 
albeit with a non-isotropic dispersion relation $\omega=|k_y|$ as it 
propagates in $y$ direction only. 
Thus, it can be quantized in the usual manner 
\bea
\label{eq:quant-Phi}
\hat\Phi_0(t,\f{r})=\int\frac{d^2k}{2\pi}\;
\frac{\hat b_{\fk{k}}\,\exp\{i\f{k}\cdot\f{r}-i|k_y|t\}}{\sqrt{2|k_y|}}
+{\rm h.c.}
\ea
with standard bosonic creation and annihilation operators 
$\hat b_{\fk{k}}^\dagger$ and $\hat b_{\fk{k}}$ 
satisfying 
$[\hat b_{\fk{k^{\phantom{\prime}}}}^{\phantom{\dagger}},\hat b_{{\fk{k'}}}^\dagger]=\delta^2(\f{k}-\f{k'})$.
They correspond to the initial vacuum state $\ket{0}_{\rm in}$ of the environment field 
(incoming from $\mathcal{I}^-$) with $\hat b_{\fk{k}}\ket{0}_{\rm in}=0$. 
In the following, we will use the notation $k_y\to\kappa$ and $k_x\to k$ for 
reasons of convenience.

\subsection{Decoupled Case}\label{sec:Decoupled Case}

\begin{figure}[htbp]
\centering
\includegraphics[width=\linewidth]{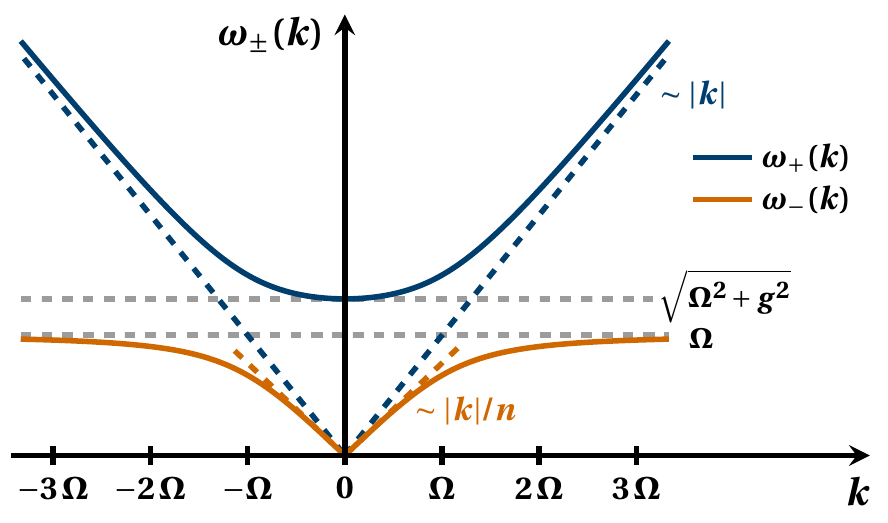}
\caption{Plot of the dispersion relation $\omega_{\pm}(k)$ in Eq.~\eqref{eq:two-bands}
with exemplary parameters $g/\Omega = 5/6$. 
}
\label{fig:dispersion}
\end{figure}

As pointed out at the beginning of this section, 
the situation is different for an initially non-dissipative 
(albeit still dispersive) medium.
In this case, the coupled $A$-$\Psi$ system is decoupled from the $\Phi$ field and 
thus both start to evolve from their independent vacuum states. 
For $G=0$ and non-vanishing, constant $\Omega$ and $g$, 
we have the usual Hopfield Hamiltonian $\hat H_{\rm H}$ 
which can be diagonalized \cite{Huttner_1991} via 
\bea
:\!\hat H_{\rm H}\!:\;
=
\sum_\pm\int dk\,\omega_\pm(k) \hat a^\dagger_\pm(k)\hat a_\pm(k) 
\,,
\ea
with $\hat a^\dagger_\pm(k)$ and $\hat a_\pm(k)$ denoting 
the creation and annihilation operators of the two bands 
\bea
\label{eq:two-bands}
\omega_\pm(k)=\sqrt{\frac{(k^2+\Omega^2+g^2)\pm\rho(k)}{2}}
\,,
\ea
where we have used the abbreviations 
\bea
\rho(k)=\sqrt{4k^2g^2+\sigma^2(k)}
\ea
and 
\bea
\sigma(k)=k^2-g^2-\Omega^2
\,.
\ea
As illustrated in Fig.~\ref{fig:dispersion},
the lower band $\omega_-(k)$ behaves as $|k|/n$ for small $k$, 
while the upper band $\omega_+(k)$ tends to a constant value 
$\sqrt{\Omega^2+g^2}=n\Omega$. 
For large $k$, on the other hand, the lower band $\omega_-(k)$ 
approaches the medium resonance frequency $\Omega$ while the 
upper band reaches the vacuum light cone $|k|$. 
In contrast to the lower band accounting for massless photons at small wave numbers $k$, 
the upper band resembles the dispersion relation for a relativistic massive field.
Note also the band gap of width $(n-1) \Omega$ between the two bands.

\section{Particle creation} 

Based on the model established above, one may study various 
quantum effects in and out of equilibrium. 
Since providing an \textit{ab initio} treatment of time-dependent 
dissipative media is one of the major benefits our approach offers 
compared to existing models, we will henceforth focus on 
non-equilibrium phenomena.
In principle, one could consider time-dependent 
parameters $\Omega(t)$, $g(t)$, or $G(t)$; or a combination of them.
In order to illustrate the novel features of our model 
(in comparison to non-dissipative Hopfield dielectrics), let us consider 
a scenario where we switch on and off dissipation by 
a time-dependent $G(t)$ with $G(t\to\pm\infty)=0$ while 
the other two parameters $\Omega$ and $g$ are kept constant. 

Even for constant $\Omega$ and $g$, solving the 
decoupled field equation~\eqref{eq:decouple-Psi} 
is quite involved for general profiles $G(t)$,
which makes it hard to reach progress analytically.  
A major difficulty arises from the interplay of excitation and dissipation, 
i.e., particles are already damped while they are created.
In order to focus on the phenomenon of particle creation 
(and to separate it from the competing damping effect),
we assume that the coupling $G(t)$ is switched on for a sufficiently 
short time and to a maximum value which is not too large, such that 
the damping during this switching time can be neglected 
in a first approximation. 

\subsection{Perturbation Theory} 

Formally, the approximation described above can be implemented via 
perturbation theory based on a power expansion in $G$.
As one option, this can be formulated in the framework of 
time-dependent perturbation theory with the perturbation 
Hamiltonian $\hat H_{\Psi\Phi}$ stemming from the 
Lagrangian $L_{\Psi\Phi}$ given in Eq.~\eqref{eq:LPhiPsi}, 
while the remaining contributions 
$L_{A} + L_{\Psi} + L_{A\Psi} + L_{\Phi}$ correspond
to the undisturbed $\hat H_0$ problem. 
As another option, we may approximate the equations of motion for 
the field operators by omitting all terms of order $\ord(G^2)$. 
Decoupling the original problems 
\eqref{eq:eulerLagrangeEquationA}, 
\eqref{eq:eulerLagrangeEquationPsi} and
\eqref{eq:eulerLagrangeEquationPhi} 
for time-dependent $G(t)$ in this way 
yields the simplified expression
\bea
\label{eq:approximate-A-problem}
\left[
\left(\partial_t^2+\Omega^2\right)
\left(\partial_t^2-\partial_x^2\right)
+g^2\partial_t^2 
\right]
\hat A(t,x)
= \qquad\quad
\nn
-g \partial_t G(t)\partial_t \hat\Phi_0(t,x,y=0)
+\ord(G^2)
\,,\quad
\ea
see also Eq.~\eqref{eq:A-from-Phi}. 
After inserting the inhomogeneity~\eqref{eq:quant-Phi}, 
comparing the initial $\hat A^{\rm in}(t,x)$ and 
final $\hat A^{\rm out}(t,x)$ solutions 
(both expressed in terms of the creation and annihilation 
operators introduced in Sec.~\ref{sec:Decoupled Case}) 
yields the Bogoliubov transformation
(to first order in $G$)
\bea
\label{eq:Bogoliubov}
\hat a_\pm^{\rm out}(k)=\hat a_\pm^{\rm in}(k)+
\int d\kappa
\left(
\alpha_{k\kappa}^\pm\hat b_{k\kappa}+
\beta_{k\kappa}^\pm\hat b_{-k\kappa}^\dagger
\right) 
\,. 
\ea
The Bogoliubov coefficients $\alpha_{k\kappa}^\pm$ and $\beta_{k\kappa}^\pm$
connecting the initial $\hat a_\pm^{\rm in}(k)$ and final $\hat a_\pm^{\rm out}(k)$ 
annihilation operators (i.e., before and after switching on an off dissipation) 
with the initial environment operators $\hat b_{k\kappa}$ and $\hat b_{k\kappa}^\dagger$
are proportional to the Fourier transform $\widetilde G(\omega)$ 
of the switching function $G(t)$, evaluated at $\omega_\pm(k)-|\kappa|$ and 
$\omega_\pm(k)+|\kappa|$, respectively, plus $\ord(G^2)$ corrections. 

To lowest order in $G$, the number (density) 
of particles created per unit length is given by 
\bea
\label{eq:photonNumber}
\langle\hat n_\pm^{\rm out}(k)\rangle_{\rm in}
&=&
\int d\kappa\,\left|\beta_{k\kappa}^\pm\right|^2
=
\frac{\rho(k) \mp \sigma(k)}{8 
\rho(k) \omega_\pm(k)}
\int d\kappa\,|\kappa|\,\times
\nn
&&
\times 
\left|\widetilde G\left(\omega_\pm(k)+|\kappa|\right)\right|^2 
\,.
\ea
Assuming that the characteristic rate of change in the switching 
function $G(t)$ is much slower than the medium frequency $\Omega$,
the number $\langle\hat n_+^{\rm out}(k)\rangle_{\rm in}$
of particles in the upper band $\omega_+(k)$ is exponentially
suppressed due to $\omega_+(k)\geq\sqrt{\Omega^2+g^2}$.
For the same reason, we may approximate the lower band 
according to $\omega_-(k)\approx|k|/n$ 
\footnote{Without this linearization, 
the total number~\eqref{eq:photonYieldCoupling} 
of particles created in the lower band 
would diverge after integrating over all $k$ 
because $\omega_-(k\to\pm\infty)=\Omega$ 
approaches a constant value at large $k$. 
However, this is just an artifact of the idealized coupling in 
Eq.~\eqref{eq:LPhiPsi} which does not have a UV cut-off.}. 

\subsection{Lorentzian Profile} 

A particularly simple expression can be obtained for a switching 
function in the form of a Lorentz pulse 
\bea
\label{eq:Lorentz}
G(t)=G_0\frac{\tau^2}{\tau^2+t^2}
\,,
\ea
with the characteristic switching time $\tau>0$.
In this case, the Fourier transform is just an exponential function 
$\widetilde G(\omega)=G_0\tau\sqrt{\pi/2}\exp\{-\tau|\omega|\}$ and the total 
number of created particles $N$ per 
length $\ell$ reads 
\bea
\label{eq:photonYieldCoupling}
\frac{N}{\ell}
\approx 
\frac{1}{32}\,\frac{g^2}{\sqrt{\Omega^2+g^2}}\,\frac{G_0^2}{\Omega^3\tau^2}
=\frac{\Gamma_0}{(\Omega\tau)^2}\,\frac{n^2-1}{8n}
\,.
\ea
Since we have assumed a slow switching function, i.e., 
$\Omega\tau\gg1$, a significant number $N$ of photons can only 
be created by switching 
the dissipation in a region of sufficiently large optical path length $n\ell$. 
Even though the above result was obtained for the specific switching 
function~\eqref{eq:Lorentz}, the qualitative scaling behavior should 
be the same for other (reasonable) profiles $G(t)$. 

Let us compare the above number to the well-known case of changing 
the refractive index $n(t)$ by a small amount $\Delta n\ll1$ in 
absence of dissipation, see, e.g., 
\cite{Liberati_2012, Belgiorno_2014, Tian_2017, Lang_2019}. 
For a Lorentzian perturbation $\Delta n(t)$ analogous to Eq. \eqref{eq:Lorentz}, 
the number of particles $N$ per unit length $\ell$ reads 
\bea
\frac{N}{\ell}
=
\frac{\pi}{16}\,\frac{(\Delta n)^2}{n\tau}
\,.
\ea
In non-linear dielectric media, refractive index perturbations $\Delta n(t)$ 
of order ${\cal O}(10^{-3})$ can be generated by strong laser pulses 
and the Kerr effect \cite{Belgiorno_2010, Faccio_2010}. 
Slightly stronger perturbations $\Delta n(t)$ 
of order ${\cal O}(10^{-2})$ have been reported for tunable 
meta-materials \cite{Laetheenmaeki_2013}. 
However, since the number of created particles $N/\ell$
is of second order in $\Delta n$, 
switching dissipation could be more effective.

\subsection{Partner Particles} 

As is well known, changing the refractive index $n(t)$ creates photons in pairs 
with opposite momenta $\pm k$.  
The relation between photons and their partners can be observed in the 
two-point correlation function $\langle\hat A(t,x)\hat A(t,x')\rangle$, 
for example. 
For times $t$ long after the switch, one 
obtains distinctive signatures at distances $|x-x'|=2t/n+\ord(\tau)$, 
see also \cite{Prain_2010, Lang_2019}. 

In contrast, the partners of photons created by switching 
on and off dissipation are not other 
medium photons, but excitations of the environment field $\Phi$.
This can already be inferred from the (lowest-order) Bogoliubov 
transformation~\eqref{eq:Bogoliubov}, 
see \cite{Hotta_2015}. 
As another signature, we find pairs of peaks in the correlation 
function $\langle\hat\Phi(t,x,y)\hat A(t,x')\rangle$ at 
distances $|y| = t + {\cal O}(\tau)$ 
and $|x-x'| = t/n + {\cal O}(\tau)$, 
but not (to first order) in the correlation $\langle\hat A(t,x)\hat A(t,x')\rangle$.   

Apart from this, there is no first-order imprint in the two-point function 
$\langle\hat \Phi(t,x,y)\hat \Phi(t,x',y')\rangle$, which indicates again that all 
excitations created in the $\Phi$ field have partners in the medium. 
Therefore, we obtain no pairs of correlated  
$\Phi$ excitations (to lowest order), in contrast 
to another mechanism of quantum radiation studied in 
Ref.~\cite{Ciuti_2005}, where both partners eventually escape 
to a surrounding field. 

\subsection{Sudden Switching} 

Previous works including Refs.~\cite{Ciuti_2005,  Liberato_2017} have 
simplified their analysis by considering scenarios in which the light-matter 
coupling is suddenly switched off. 
This simplification is not necessary in our approach, which allows us to 
take into account the dependence on the temporal switching function $G(t)$.
For a step-like profile $G(t) = G_0 \Theta(-t)$, our perturbative 
result~\eqref{eq:photonNumber} yields divergent particle numbers 
$\langle\hat n_\pm^{\rm out}(k)\rangle_{\rm in}$ for all modes $k$. 
This singularity is caused by an ultra-violet (UV) divergence of the 
$\kappa$-integration and stems from the idealized interaction term 
$L_{\Psi \Phi}$ in our model Lagrangian~\eqref{eq:classicalLagrangianSum} 
which has no UV cutoff and thus couples each mode $k$ of the medium 
to arbitrarily large wave numbers $\kappa$ of the environment $\Phi$.
By analytically solving Eq.~\eqref{eq:decouple-Psi} in case of 
$G(t) = G_0 \Theta(-t)$ with constant $\Omega$ and $g$, we have 
found this result to apply even beyond the scope of perturbation theory.

\section{Conclusions} 

We generalized the well-known Hopfield model involving the electromagnetic
field $A$ and the medium polarization field $\Psi$ by adding an environment 
field $\Phi$. 
In this way, we arrived at a microscopic Lagrangian corresponding to a 
1+1 dimensional dielectric medium including dispersion and dissipation. 
The model is constructed in such a way that it allows for the derivation 
of quantum electrodynamics in such media without ambiguities and without 
resorting to additional assumptions 
such as the Markov approximation. 
Consequently, it naturally accounts for the dynamics in media with time-dependent 
backgrounds, which is a major benefit in comparison to existing models
for dissipative dielectrics. 

As an exemplary configuration with non-constant parameters, 
we considered switching on and off dissipation and 
derived the number of created photons in dependence on the temporal 
switching function $G(t)$ and the switching time $\tau$. 
To further illustrate the photon yield calculated above, 
let us compare two scenarios:
In scenario~I, we consider a Lorentzian pulse $G(t)$ 
of height $G_0$ and width $\tau$
within a time-dependent waveguide of length $\ell$.
In scenario~II, we envision a static waveguide of the same length $\ell$ 
with constant coupling $G_0$ (see Sec.~\ref{sec:Dispersion}). 
Now, if $\ell$ (for a given $G_0$) was sufficiently large that typical 
photons of frequencies $\omega =\ord(1/\tau)$ would be damped away 
according to Eqs.~\eqref{eq:dampedA-solution} 
and~\eqref{eq:damping-strength} before 
fully traversing the static waveguide in scenario~II, 
the corresponding scenario~I (with the same $G_0$)
would yield a particle number of order unity. 
As we switch dissipation just briefly to the strength $G_0$, 
most particles created by the modulation $G(t)$ are not dissipated but should, 
in principle, be observable after dissipation has been switched off again.
Thus, the photon yield of a short pulse $G(t)$ could exceed the quantum 
radiation generated by a time-dependent refractive index $n(t)$,
because variations $\Delta n(t)$ are typically small 
and yield photon numbers quadratic in $\Delta n$.  

Since quantum radiation typically creates particles in pairs 
(i.e., a squeezed state), another interesting question concerns 
the partner particles of the produced photons. 
In contrast to the case of a time-dependent refractive index $n(t)$ 
and other scenarios (see, e.g., \cite{Ciuti_2005, Liberato_2007, Auer_2012}), 
we find that the partner
particles of photons created by switching on and off dissipation
are (primarily) excitations of the environment field $\Phi$ 
instead of other photons.  
%

\acknowledgments 

R.S.~acknowledges stimulating exchange with Flavien Gyger.
R.S. and S.L. were supported by  
German Research Foundation, 
Grant No. 278162697 (SFB 1242). 
W.G.U.~acknowledges support from the Helmholtz Association, the Humboldt Foundation, 
the Canadian Institute for Advanced Research (CIfAR), 
the  Natural Science and Engineering Research Council of Canada, 
and the Hagler Institute for Advanced Research at Texas A\&M University.


\bibliography{DissDiel-PaperRefs} 

\end{document}